# A Gibbs Sampling Alternative to Reversible Jump MCMC

Stephen G. Walker [1]

ABSTRACT. This note presents a simple and elegant sampler which could be used as an alternative to the reversible jump MCMC methodology.

**1. Introduction.** This note is about the problem of performing posterior Bayesian inference via Markov chains when the dimension of the model is not fixed. The standard solution is the reversible jump MCMC approach described in Green (1995). The set–up is typically of the form $p(y|\theta^{(k)}, k)$, so that a model for data $y$ is described for each $k = \{1, 2, \ldots\}$ and the parameter for model of dimension $k$ is $\theta^{(k)}$. A prior distribution is now assigned to $\theta^{(k)}$, say $\pi_k(\theta^{(k)})$, and a prior for $k$, say $\pi(k)$.

So let us write the (incomplete) joint density for $(y, \theta^{(k)}, k)$ as

$$p(y, \theta^{(k)}, k) = p(y|\theta^{(k)}, k)\, \pi_k(\theta^{(k)})\, \pi(k).$$

This is incomplete (though obviously a valid model) since there is nothing about the $(\theta^{(j)}; j \neq k)$. So let us add a distribution for $(\theta^{(j)}; j \neq k)$ when model $k$ is conditioned on;

$$p(y, (\theta^{(j)}; j = 1, 2, \ldots), k) = p(y, \theta^{(k)}, k) \prod_{l=k-1}^{1} p(\theta^{(l)}|\theta^{(l+1)}) \prod_{l=k+1}^{\infty} p(\theta^{(l)}|\theta^{(l-1)}),$$

where the choice of $p(\theta^{(l)}|\theta^{(l-1)})$ and $p(\theta^{(l)}|\theta^{(l+1)})$ is arbitrary. The marginal model is correct, just integrate out the $(\theta^{(j)}; j \neq k)$. So the latent variables $(\theta^{(j)}; j \neq k)$, conditioned on $k$, would at first sight not to be needed, but they play a crucial role in that they serve as the proposal move between dimensions.

The problem now is how to move between dimensions, since the choice is infinite and so the precise probabilities can not be found. However, for this we will introduce another latent variable $u$ which facilitates the moves in that it makes the choice finite and hence probabilities can be computed. We write $p(u|k)$ for this and for simplicity of exposition, though it is easily allowed to be more general, we take $u = k+1$ with probability $q$ and $u = k$ with probability $1-q$, for all $k$. So, overall, the joint density being considered is

$$p(u, y, (\theta^{(j)}; j = 1, 2 \ldots), k) = p(u|k)\, p(y, (\theta^{(j)}; j = 1, 2, \ldots), k).$$

---

[1]Stephen G. Walker is Professor of Statistics, Institute of Mathematics, Statistics & Actuarial Science, University of Kent, Canterbury, U. K. (email: S.G.Walker@kent.ac.uk)

We are now in a position to describe the Gibbs sampler for dimension jumping.

**2. The Gibbs sampler.** Suppose the chain is currently at $k$. Then we sample $\theta^{(k)}$ from $\pi_k(\theta^{(k)}|y, k) \propto p(y|\theta^{(k)}, k)\, \pi_k(\theta^{(k)})$ in the usual way and typically this is not an issue in such dimension varying models since it is done assuming $k$ is fixed. Now, given $k$, we will also sample $\theta^{(k+1)}$ from $p(\theta^{(k+1)}|\theta^{(k)})$ and $\theta^{(k-1)}$ from $p(\theta^{(k-1)}|\theta^{(k)})$. We need these two since the moves from $k$ can be to $\{k-1, k, k+1\}$. The choice of these conditional densities is precisely for the same reasons that particular moves are suggested in the reversible jump MCMC approach; basically, to increase the chance of a move away from $k$.

Once this has been done, the $u$ is sampled, and is either $k$ or $k+1$. Let us assume it is $k+1$, so that the next $k$, we will call it $j$, can be either $k$ or $k+1$. Now, clearly, we have

$$\pi(j = k+1|u = k+1, \ldots) \propto (1-q)\, p(y, \theta^{(k+1)}, k+1)\, p(\theta^{(k)}|\theta^{(k+1)})$$

and

$$\pi(j = k|u = k+1, \ldots) \propto q\, p(y, \theta^{(k)}, k)\, p(\theta^{(k+1)}|\theta^{(k)}). \qquad (1)$$

All the other latent variables and their densities are common to, and hence cancel out from, both terms, and so are not needed.

On the other hand, if $u = k$ then $j$ can be either $k$ or $k-1$. It is then easy to derive that

$$\pi(j = k|u = k, \ldots) \propto (1-q)\, p(y, \theta^{(k)}, k)\, p(\theta^{(k-1)}|\theta^{(k)})$$

and

$$\pi(j = k-1|u = k, \ldots) \propto q\, p(y, \theta^{(k-1)}, k-1)\, p(\theta^{(k)}|\theta^{(k-1)}). \qquad (2)$$

Either way, $j$ is easily sampled.

There is a special case which deserves attention and this is when we insist on

$$p(\theta^{(k)}|\theta^{(k-1)})\, \pi_{k-1}(\theta^{(k-1)}) = p(\theta^{(k-1)}|\theta^{(k)})\, \pi_k(\theta^{(k)}) \qquad (3)$$

for all $k$. Then, for probabilities (1), we now have the simpler situation,

$$\pi(j = k+1|u = k+1, \ldots) \propto (1-q)\, p(y|\theta^{(k+1)}, k+1)\, \pi(k+1)$$

and
$$\pi(j = k | u = k+1, \ldots) \propto q\, p(y|\theta^{(k)}, k)\, \pi(k),$$
and for probabilities (2), we have
$$\pi(j = k | u = k, \ldots) \propto (1-q)\, p(y|\theta^{(k)}, k)\, \pi(k)$$
and
$$\pi(j = k-1 | u = k, \ldots) \propto q\, p(y|\theta^{(k-1)}, k-1)\, \pi(k-1).$$

Once the idea of the Gibbs sampler has been understood, a generalization to more types of moves is quite straightforward, and this would involve a modification of $p(u|k)$.

**3. Discussion.** In summary, the algorithm is as easy as follows, given $k$:

1. Sample $\theta^{(k)}$ from $\pi_k(\theta^{(k)}|y, k)$, and sample $\theta^{(k+1)}$ and $\theta^{(k-1)}$ from "proposals" $p(\theta^{(k+1)}|\theta^{(k)})$ and $p(\theta^{(k-1)}|\theta^{(k)})$, respectively.

2. Sample $u$ from $p(u|k)$ so that $u = k+1$ with probability $q$ and $u = k$ with probability $q - 1$.

3. Sample the new $j$ from the appropriate probabilities (1) or (2).

Unlike the reversible jump MCMC approach we don't actually need any special relation between $p(\theta^{(k)}|\theta^{(k-1)})$ and $p(\theta^{(k-1)}|\theta^{(k)})$, though the probabilities (1) and (2) are simplified if they do satisfy a particular relation (3). This is because there is no pressure to force a detailed balance criterion. So, we have described a Gibbs sampler version of the reversible jump MCMC approach which shares features such as the evidence of proposal moves but, as we have just said, removes some of the pressure, and also lacks the need for a Jacobian. Also, unlike reversible jump MCMC there is no need to explain the algorithm; it is self evident and remarkably simple.

It would be quite easy to demonstrate in particular cases a more efficient algorithm has been introduced when compared to the reversible jump MCMC. No doubt it would also be achievable the other way round. This is rather beside the point. What is clear is that a vastly simpler algorithm has been presented and it is clear that there is no obvious reason why it should be uniformly worse than reversible jump MCMC.

**Reference.**

Green, P. J. (1995). Reversible jump Markov chain Monte Carlo computation and Bayesian model determination. *Biometrika* **82**, 711–732.